\documentclass[aps,print,prb,twocolumn,superscriptaddress]{revtex4-2}

\usepackage[hidelinks]{hyperref}
\hypersetup{
	colorlinks,
	linkcolor={red},
	citecolor={blue},
	urlcolor={blue}
}
\usepackage{physics}
\usepackage{balance}
\usepackage{amssymb}
\usepackage{suffix}
\usepackage{mathtools}
\usepackage[utf8]{inputenc}
\usepackage{booktabs}
\usepackage{cases}
\usepackage[multiple]{footmisc}
\usepackage{dcolumn}
\usepackage{color,soul}
\usepackage{rotating}
\usepackage{perpage}
\usepackage{siunitx}
\usepackage{xcolor}
\usepackage{soul}
\usepackage{amsmath}
\usepackage{tikz}
\usepackage[T1]{fontenc}
\usepackage{etoolbox}
\usepackage{graphics}
\usepackage{siunitx}
\usepackage{float}	
\usepackage{collref}
\usepackage{multirow}
\usepackage{mathtools}
\usepackage{bm}
\usepackage{url}

\usepackage{tikz}
\usepackage{tikz-3dplot}
\usepackage{accents}

\makeatletter
\newcommand{\doublewidetilde}[1]{{%
		\mathpalette\double@widetilde{#1}%
}}
\newcommand{\double@widetilde}[2]{%
	\sbox\z@{$\m@th#1\widetilde{#2}$}%
	\ht\z@=.9\ht\z@
	\widetilde{\box\z@}%
}
\makeatother

\usepackage{etoolbox,lipsum}

\begin{document} 
	\title{Dynamical phase transition in a strongly hybridized phonon-triplon chain}
	
	\author{Mohsen Yarmohammadi}
	\email{mohsen.yarmohammadi@georgetown.edu}
	\address{Department of Physics, Georgetown University, Washington DC 20057, USA} 
	\date{\today}
	\begin{abstract}
		We study a dimerized spin-1/2 chain, such as \(\mathrm{CuGeO}_3\), hosting triplon excitations coupled to optical phonons under weak terahertz laser driving. Both phonons and triplons weakly lose energy into the surrounding baths, forming a non-equilibrium steady state. In the strong phonon-triplon coupling regime, phonons near the two-triplon continuum hybridize strongly with triplons. Using mean-field Lindblad dynamics, we show that this strong hybridization induces sharp first-order phase transitions—either single or simultaneous double—in the emission spectrum, mainly due to dissipation-induced nonlinearities. Using mean-field Floquet analysis of harmonic modes in both sectors, we analytically confirm the existence of these phase transitions. Furthermore, we map the complete steady-state phase diagram by varying key control parameters and provide experimentally relevant parameters for observing these transitions in laser-driven \(\mathrm{CuGeO}_3\).
	\end{abstract}
	
	\maketitle
	{\allowdisplaybreaks
		
		\section{Introduction}
		The advent of ultrafast laser techniques for driving quantum materials out of equilibrium has opened exciting avenues for exploring and controlling emergent phenomena on ultrashort timescales~\cite{delatorre_nonthermal_2021,Bao2022,Buzzi2018,Mitrano2020,Xu2025}. In this context, the study of nonequilibrium steady states (NESS) has emerged as a powerful framework for understanding the interplay between external driving and dissipation in quantum materials~\cite{ikeda_general_2020,murakami_nonequilibrium_2017,aydogan_parameter_2025,PhysRevResearch.6.023160}. While transient dynamics decay toward equilibrium, NESS arise from a sustained balance between drive and dissipation. They host rich phenomena, including dynamical phase transitions, emergent steady-state order, and externally tunable quantum states~\cite{PhysRevResearch.4.023115,doi:10.1126/sciadv.aao0043,PhysRevResearch.6.033075,doi:10.1126/sciadv.aax1568,10.1093/ptep/ptad007}.
		
		Among ultrafast control mechanisms, magnetophononics—where optically excited phonons dynamically couple to magnetic degrees of freedom—has emerged as a powerful route for manipulating quantum magnets~\cite{fechner_magnetophononics_2018,afanasiev_ultrafast_2021,PhysRevB.110.064420, PhysRevB.107.174415,PhysRevB.103.045132,PhysRevB.107.184440}. This approach enables the coherent modulation of magnetic exchange, offering promising pathways toward novel functionalities in quantum technologies. Accordingly, systems with strong spin-phonon coupling (SPC) are particularly compelling, as lattice vibrations can significantly reshape magnetic excitations and influence the system’s behavior in nontrivial ways~\cite{https://doi.org/10.1002/apxr.202300153,Xu2022,PhysRevB.107.075421,doi:10.1126/sciadv.aar5164}.
		
		Dimerized spin chains are a well-studied class of gapped quantum magnets, valued for their well-defined excitation spectra and tunable properties under external perturbations~\cite{yu_excitation_1999,taddel_finite-temperature_2018,Zhang2024,he_dimerization_vectorchirality_2024,CAO2024415876,BRENIG19981808,PhysRevB.61.6126,PhysRevB.110.L201112,PhysRevResearch.6.L032050}. These systems host triplons—spin-1 quasiparticles originating from singlet-triplet transitions on dimers—with a finite energy gap that enables selective, resonant driving of phonon modes coupled to the triplons. Recent studies have shown that such driven phonons can give rise to rich nonequilibrium phenomena, including coherent spin dynamics, phonon-mediated interactions, and high-harmonic generation~\cite{PhysRevB.95.174407,Forst2011,PhysRevB.110.064420,PhysRevB.107.184440,PhysRevB.107.174415,PhysRevB.103.045132,doi:10.1073/pnas.2204219119,Mrudul2020}. Despite growing interest, the role of strong feedback between hybridized spins and driven phonons in shaping NESS remains largely unexplored. We ask: How do NESS evolve in strongly hybridized spin-phonon systems in driven-dissipative quantum magnets? Can distinct NESS emerge in this regime, and do they exhibit new dynamical phases?

		In this work, we address these open questions by studying a dimerized spin-1/2 chain inspired by \(\mathrm{CuGeO}_3\)~\cite{PhysRevB.63.094401,Chen2021,VANLOOSDRECHT19971017,BUCHNER1999956,YUASA20081087,spitz2025phononspectrumspinpeierlsphase}, subjected to a continuous-wave terahertz~(THz) laser. The laser selectively excites optical phonons coupled to triplons, enabling controlled exploration of nonequilibrium lattice-spin interplay. Both phonons and triplons dissipate into thermal baths, forming a rich driven-dissipative setting where coherent drive, strong SPC, and dissipation interact. Using a combination of bond-operator formalism and mean-field Lindblad dynamics, we show that strong SPC near the two-triplon continuum leads to first-order phase transitions in the emission powers. These transitions arise from strong hybridization- and dissipation-induced nonlinearities.
		
		This behavior resembles transitions observed in other driven-dissipative systems~\cite{Heyl_2018,PhysRevLett.113.265702,PhysRevA.86.012116,PhysRevA.98.042118,PhysRevLett.110.233601,PhysRevA.110.042209,PhysRevA.97.013853,Walker2018,Fink2018,Benary_2022,PhysRevB.110.104301,PhysRevE.110.064156,PhysRevResearch.5.033178,Tang_2022,PhysRevLett.115.140602}. However, unlike conventional dynamical phase transitions—such as Loschmidt singularities in time, Fisher zeros, or scaling behavior in relaxation dynamics—this transition occurs in the NESS~\cite{PhysRevB.108.L140305,PhysRevB.111.L201116,PhysRevB.109.224417} and only arises when the phonon energy is near the two-triplon continuum and the SPC is strong—conditions absent in earlier studies. Using mean-field Floquet analysis in the NESS, we analytically confirm the phase transitions and map the full phase diagram across key parameters. To our knowledge, such a comprehensive analysis has not been previously performed for dimerized spin chains strongly hybridized with phonons. Two videos illustrating the presence or absence of the dynamical phase transitions at various regimes are also provided in the Supplemental Material~(SM)~\cite{SM}.
		
		The paper is organized as follows. In Sec.~\ref{S2}, we introduce the system and its coupling to a bath. Section~\ref{S3} presents the main results, including a Floquet analysis of the steady states and associated phase transitions, along with an experimental perspective. Finally, Sec.~\ref{S4} summarizes our findings and briefly outlines future directions.
		
		\section{Theoretical background}\label{S2}
		We consider a dimerized spin-\(1/2\) chain, motivated by materials such as \(\mathrm{CuGeO}_3\)~\cite{PhysRevB.63.094401,Chen2021,VANLOOSDRECHT19971017,BUCHNER1999956,YUASA20081087}, and study its response under the influence of a THz laser field. The compound \(\mathrm{CuGeO}_3\) serves as a prototypical example of a quasi-one-dimensional spin-\(1/2\) antiferromagnet, exhibiting a spin-Peierls transition at low temperatures. Below the critical temperature of approximately \(14\,\text{K}\), the system undergoes spontaneous lattice dimerization, driven by the coupling between spin and lattice degrees of freedom. This dimerization doubles the unit cell and stabilizes a nonmagnetic singlet ground state, characterized by a finite spin gap and suppressed magnetic susceptibility.
		
		As depicted in Fig.~\ref{f1}, the Cu-O chains oriented along a specific crystallographic direction host spin-\(1/2\) moments from Cu ions, which interact via alternating intradimer (\(J\)) and interdimer (\(J' < J\)) exchange couplings. These exchange interactions are sensitive to lattice distortions mediated by the vibrations of intermediate oxygen~(O) atoms. The electric field component of the laser can drive these lattice vibrations, thereby modulating the exchange pathways. This magnetophononic coupling mechanism enables indirect laser control of spin dynamics through SPC~(see Sec.~\ref{S2A}), while energy dissipation is incorporated via a phenomenological thermal bath in Sec.~\ref{S2B}.
		
		\subsection{Hamiltonian model and drive}\label{S2A}
		We study a time-periodically driven spin-phonon system described by a Hamiltonian that encapsulates the key elements of the hybridized phonon-triplon dynamics under external optical excitation. For simplicity, we set \(\hbar = 1\) and denote operators by their symbols without hats, i.e., \(\hat{O} = O\). The total Hamiltonian of the close Cu-O chain is given by
		\begin{equation}
			H = H_s + H_p + H_{lp} + H_{sp}\,,
		\end{equation}
		where \(H_s\) describes the spin chain, \(H_p\) corresponds to the phonon sector, \(H_{lp}\) represents the laser-phonon coupling, and \(H_{sp}\) accounts for the spin-phonon interaction.
		\begin{figure}[t]
			\centering
			\includegraphics[width=0.9\linewidth]{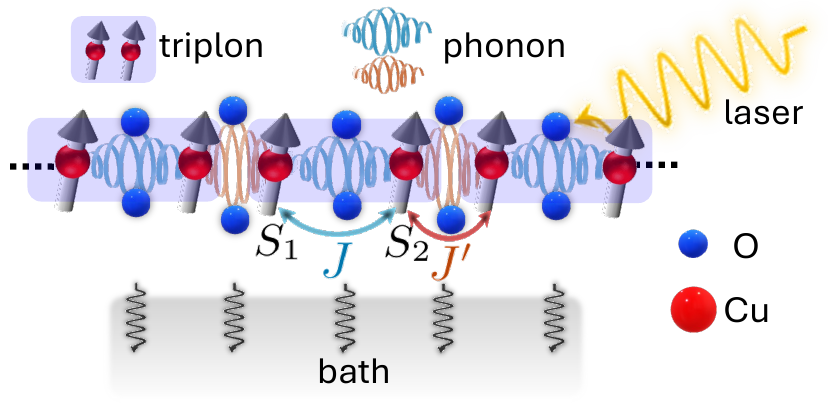}
			\caption{Schematic illustration of a driven-dissipative phonon-triplon chain, exemplified by \(\mathrm{CuGeO}_3\). The spin chain exhibits dimerization, with alternating spins labeled \(S_1\) and \(S_2\) to denote the two inequivalent sites. Triplon excitations (blue dimers) and exchange couplings \(J\) and \(J'\) are modulated by optical phonons (red and blue springs). An ultrafast terahertz laser field selectively drives infrared-active optical phonon modes, inducing nonequilibrium spin dynamics (see Sec.~\ref{S2A}). Coupling to an external bath (Sec.~\ref{S2B}) facilitates energy dissipation and allows the system to reach a steady state.}
			\label{f1}
		\end{figure}
		
		The spin Hamiltonian for the dimerized chain is given by\begin{equation}
			H_s = J \sum^L_{\ell = 1} \left(\vec{S}_{1,\ell} \cdot \vec{S}_{2,\ell} + J' \vec{S}_{2,\ell} \cdot \vec{S}_{1,\ell+1}\right)\,,
		\end{equation}with \( J'/J = 1/2 \). Here, \( \vec{S}_{1,\ell} \) and \( \vec{S}_{2,\ell} \) denote spin-\(1/2\) operators corresponding to the two spins in the \( \ell \)-th dimer of a chain with \( L \) dimers, as illustrated in Fig.~\ref{f1}.
		
		The phonon system is modeled as a single harmonic mode near the center of the phonon dispersion, which predominantly couples to long-wavelength THz laser fields~\cite{PhysRevB.95.174407,Forst2011,PhysRevB.110.064420,PhysRevB.107.184440}. The corresponding Hamiltonian is\begin{equation}
			H_p = \omega_0 a^\dagger a\,,
		\end{equation}where \( a^\dagger \) and \( a \) are the phonon creation and annihilation operators, and \( \omega_0 \) is the phonon frequency. Since the phonon couples to the laser’s electric field, it must be an infrared-active optical mode. All other phonons are treated as main part of the surrounding thermal bath.
		
		The laser field couples to the phonon via a classical monochromatic electric field \( E(t) = \mathcal{E} \cos(\Omega t) \), where \( \mathcal{E} \) and \( \Omega \) are the field amplitude and frequency, respectively. This coupling is described by\begin{equation}
			H_{lp} = \sqrt{L} E(t) (a + a^\dagger)\,,
		\end{equation}representing the interaction between the phonon's dipole moment and the laser field.
		
		The leading SPC is linear in the phonon displacement, defined as \( x = (a + a^\dagger)/\sqrt{L} \), and affects both intradimer and interdimer interactions:\begin{equation}\label{eq_5}
			H_{\text{sp}} = x \sum^L_{\ell = 1}  \left( g \vec{S}_{1,\ell} \cdot \vec{S}_{2,\ell} + g'  \vec{S}_{2,\ell} \cdot \vec{S}_{1,\ell+1} \right),
		\end{equation}where \( g \) and \( g' \) are the coupling strengths to the intradimer and interdimer bonds, respectively. 
		
		To analyze the spin dynamics, we employ the bond-operator formalism~\cite{sachdev_bond-operator_1990,gopalan_spin-ladders_1994,PhysRevB.82.054404}, representing the spin operators in terms of bosonic triplons (\( t^\dagger \), \( t \)). The singlet state is treated as the vacuum, and triplet excitations are considered. The spin operators for the \( \ell \)-th dimer are expressed as\begin{subequations}\begin{align}
			S^\alpha_{1,\ell} &= \frac{1}{2} \bigg( t_{\alpha,\ell} + t^\dagger_{\alpha,\ell} - i \sum_{\beta,\zeta} \epsilon_{\alpha\beta\zeta} t^\dagger_{\beta,\ell} t_{\zeta,\ell} \bigg)\,, \\
			S^\alpha_{2,\ell} &= \frac{1}{2} \bigg( -t_{\alpha,\ell} - t^\dagger_{\alpha,\ell} - i \sum_{\beta,\zeta} \epsilon_{\alpha\beta\zeta} t^\dagger_{\beta,\ell} t_{\zeta,\ell} \bigg)\,,
		\end{align}\end{subequations}where \( \epsilon_{\alpha\beta\zeta} \) is the Levi-Civita tensor, and \( \alpha = \{x, y, z\} \) denotes the spin components.
		
		Since the model describes the spin-Peierls phase at low temperatures, triplon-triplon interactions are neglected due to their low occupancy. Using the Fourier and Bogoliubov transformations, $u_{k,\alpha} = t_{k,\alpha} \cosh(\theta_k) + t^\dagger_{-k,\alpha} \sinh(\theta_k)$, the spin Hamiltonian is diagonalized in momentum space, yielding~\cite{PhysRevB.110.064420, PhysRevB.107.174415,PhysRevB.103.045132}\begin{equation}\label{eq_8}
			H_s \approx H_0 + \sum_{k,\alpha} \varepsilon_k \, t^\dagger_{k,\alpha} t_{k,\alpha}\,,
		\end{equation}where the ground-state energy is given by \( H_0 = -\frac{3}{4} \left[ L J + J' \sum_k \cos(k) \right] \), with triplon dispersion \( \varepsilon_k = J \sqrt{1 - (J'/J) \cos(k)} \), and the Bogoliubov angle defined via \( e^{-2\theta_k} = \varepsilon_k / J \).
		
		To minimize finite-size effects in simulations, we consider a sufficiently long chain. Since the phonon number scales with the chain length \( L \), quantum fluctuations—which scale as \( 1/\sqrt{L} \)—become negligible for large \( L \), allowing us to approximate the spin-phonon interaction using mean-field decoupling, yielding~\cite{PhysRevB.110.064420, PhysRevB.107.174415,PhysRevB.103.045132}\begin{equation}
			H_{sp} \approx x \sum_{k,\alpha} \left(A_k\, t^\dagger_{k,\alpha} t_{k,\alpha} + \tfrac{B_k}{2} \big( t^\dagger_{k,\alpha} t^\dagger_{k,\alpha} + \text{h.c.} \big) \right),
		\end{equation}with matrix elements $A_k = \frac{g J}{\varepsilon_k} \left(1 - \frac{J'}{2J} \cos(k)\right) - \frac{g' J}{2 \varepsilon_k} \cos(k)$ and $B_k = \frac{(g J' - g' J)}{2 \varepsilon_k} \cos(k)$.
		
		\subsection{Dissipation and equations of motion}\label{S2B}
		To incorporate dissipation, we employ a Markovian Lindblad master equation to model the system's dynamics in contact with a thermal bath~\cite{Lindblad1976}. The time evolution of an observable \( O \) is given by{\small \begin{equation}
			\langle \dot{O} \rangle(t) = i \langle [H, O] \rangle(t) + \sum_i \gamma_i \left( \langle L_i^\dagger O L_i \rangle - \tfrac{1}{2} \langle \{ L_i^\dagger L_i, O \} \rangle \right)(t),
		\end{equation}}where the jump operators are \( L_i = \{ t_{k,\alpha}, t^\dagger_{k,\alpha}, a, a^\dagger \} \) in our system, and the corresponding damping rates are $\gamma_i = \{ \gamma_s [n(\varepsilon_k) + 1], \gamma_s n(\varepsilon_k), \gamma_p [n(\omega_0) + 1], \gamma_p n(\omega_0) \}$. Here, \( n(\varepsilon_k) \) and \( n(\omega_0) \) denote the bosonic occupation numbers of triplon and phonon modes, respectively. In the low-temperature limit (\( T \to 0 \)), where the chain is strongly dimerized, these occupation numbers vanish, and the dominant dissipative processes are emission. Consequently, the relevant jump operators reduce to \( L_i = \{ t_{k,\alpha}, a \} \), with damping rates \( \gamma_i = \{ \gamma_s, \gamma_p \} \).
		
		Ultimately, the mean-field dynamics are governed by coupled equations for the phonon displacement $q = \langle x \rangle$, momentum $p = i\langle (a^\dagger - a)/\sqrt{L}\rangle$, and triplon observables $n_k = \sum_\alpha \langle t^\dagger_{k,\alpha} t_{k,\alpha} \rangle$, and $z_k = \langle t^\dagger_{k,\alpha} t^\dagger_{k,\alpha} \rangle$:
		{\small \begin{subequations}
				\begin{align}
					\dot{q}(t) &={} \omega_0 \,p(t) - \tfrac{\gamma_p}{2}\, q(t), \label{eq_11a}\\
					\dot{p}(t) &= {}-\omega_0\, q(t) - 2\, \widetilde{E}(t) - \tfrac{\gamma_p}{2}\, p(t), \label{eq_11b}\\
					\dot{n}_k(t) &= {}2\, B_k\, q(t)\, \imaginary[z_k(t)] - \gamma_s\, n_k(t), \label{eq_11c}\\
					\dot{z}_k(t) &= 2\,i\left( \varepsilon_k + A_k q(t) \right) + 2i B_k\,q(t) \left( n_k(t) + \tfrac{3}{2} \right) -\gamma_s\, z_k(t) \,,\label{eq_11d}
		\end{align}\end{subequations}}where $\widetilde{E}(t) = E(t) + (1/L) \sum_k ( A_k n_k(t) + B_k \real[z_k(t)])$ is the effective laser field, incorporating feedback from SPC.
		
		The equations of motion enable a straightforward calculation of emitted powers in different sectors. In this work, we focus on the triplon emission power, \(\mathcal{P}(t)\), as the primary observable. Although the phonon emission power can be similarly obtained from the equation of motion for the phonon occupation \( n_p = \langle a^\dagger a \rangle / L \), it does not provide additional qualitative insight beyond what is captured by \(\mathcal{P}(t)\) for our purposes. Notably, using the relation \(\mathcal{P}(t) \sim \langle \dot{H}_s \rangle\) for the triplon sector from Eq.~\eqref{eq_8}, and incorporating the dissipative term from Eq.~\eqref{eq_11c}, we calculate the triplon emission power per unit area in \(\mathrm{CuGeO}_3\), taking into account the molar density \(\rho\) and sample thickness \(d\). The total mean-field triplon emission power per spin is thus given by\begin{align}\label{eq_12}
			\mathcal{P}(t) = \frac{\rho \, d \, n_a \, \gamma_s}{2L} \sum_k \varepsilon_k n_k(t) \,,
		\end{align}where \( n_a \) is Avogadro’s number. Throughout the manuscript, we normalize the power using the reference scale \(\mathcal{P}_0 = \rho \, d \, n_a J^2 / 2L\).

		\section{Results and discussion}\label{S3}
		We simulate the time evolution of the mean-field emission power for a chain of length \(L = 1000\) to determine if the system attains a genuine NESS. We focus on the strong SPC regime, \( g/J > g'/J' \) with \( J'/J < 1 \), to ensure that lattice vibrations have a significant impact on magnetic excitations. Later, a scan over SPC strengths identifies the coupling range where nontrivial behavior emerges. To avoid thermal melting, we use weak laser fluence for controlled energy absorption. Damping is treated via weak system-bath coupling, as the Markovian Lindblad formalism breaks down at strong damping. The coupled nonlinear equations are solved numerically with MATLAB’s \texttt{ode45} adaptive Runge-Kutta solver, balancing accuracy and efficiency. Convergence is checked by varying solver tolerances and step sizes. We set \(J=1\) (about 10 meV \(\approx\) 2.42 THz in \(\mathrm{CuGeO}_3\)) and choose optical phonon frequencies near the two-triplon continuum to guarantee sufficiently strong phonon-triplon hybridization capable of inducing a distinct phase in the NESS.
		
		\subsection{Numerical mean-field dynamical first-order phase transition in the NESS}
		To investigate the system's dynamics, we evaluate the triplon emission power \(\mathcal{P}(t)\) over time for various driving frequencies \(\Omega/J\). The results in Fig.~\ref{f2}(a)--(c) correspond to increasing phonon frequencies \(\omega_0/J = 1.25\), \(1.3\), and \(1.35\), respectively. For \(J'/J = 1/2\), the triplon dispersion \(\varepsilon_k = J \sqrt{1 - (J'/J) \cos(k)}\) produces a one-triplon band spanning energies from \(\sqrt{2}J/2\) to \(\sqrt{6}J/2\), and a two-triplon continuum from \(\sqrt{2} J\) to \(\sqrt{6} J\). The chosen phonon frequencies lie within the gap between these bands. In the weak SPC regime, resonances typically satisfy \(\Omega = \omega_0 = n \varepsilon_k\) with \(n = 1, 2\). Between the one- and two-triplon resonances, the laser and phonon fields couple more strongly near the two-triplon band due to the SPC structure in Eq.~\eqref{eq_5}, where phonons interact with both spins in a dimer rather than a single spin. This mechanism naturally enhances two-triplon over one-triplon processes. In the strong SPC regime relevant here, with \(g/J = 0.5\) and \(g'/J' = 0.3\), phonons and triplons become strongly renormalized via mutual dressing. This hybridization modifies the excitation spectrum, shifting the pronounced emission power response slightly away from \(\Omega = \omega_0\), as shown in Fig.~\ref{f2} for all three phonon frequencies.\begin{figure}[t]
			\centering
			\includegraphics[width=0.9\linewidth]{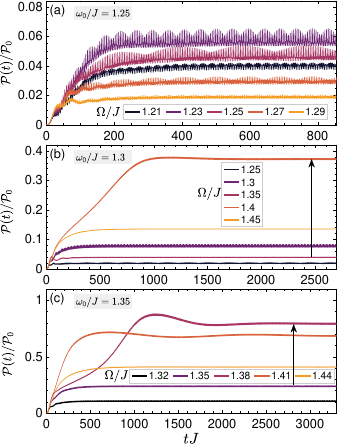}
			\caption{Mean-field time evolution of the triplon emission power, \(\mathcal{P}(t)\), across a range of driving frequencies \(\Omega/J\), shown for different phonon frequencies \(\omega_0/J\). Parameters: \(g/J = 0.5\), \(g'/J' = 0.3\), \(\mathcal{E}/\omega_0 = 0.01\), \(\gamma_p/\omega_0 = 0.05\), and \(\gamma_s/J = 0.01\). 
				(a) For \(\omega_0/J = 1.25\), the system reaches a low-power steady state with persistent oscillations. 
				(b) At \(\omega_0/J = 1.3\), a sharp jump in steady-state power appears near \(\Omega/J \approx 1.4\) (indicated by the arrow), signaling a dynamical first-order phase transition (see main text). 
				(c) For \(\omega_0/J = 1.35\), the system settles into a high-power steady state near \(\Omega/J \approx 1.38\), again signaling the emergence of a dynamically distinct phase beyond the transient regime.}
			\label{f2}
		\end{figure}
		
		In Fig.~\ref{f2}(a), for \(\omega_0/J = 1.25\), the system remains in a low-emission regime across the full range of driving frequencies \(\Omega/J\) considered. The emission power exhibits large-amplitude oscillations atop a transient response before settling into a steady-state plateau. The absence of notable changes in either the transient dynamics or the NESS indicates that the phonon frequency is too far detuned from the two-triplon band to induce a collective phonon-triplon response. As discussed below, the emission amplitude is highly sensitive to whether the phonon lies within or near the two-triplon continuum.
		
		Figure~\ref{f2}(b) shows a qualitatively different behavior as the phonon frequency is increased to \(\omega_0/J = 1.3\). At first glance, the emission amplitude is clearly reduced compared to the previous case. At a critical driving frequency \(\Omega_c/J \approx 1.4\), marked by the upward arrow, the long-time emission power exhibits a sharp, discontinuous jump (see below for confirmation of this discontinuity). This sudden change characterizes a dynamical first-order phase transition, separating two distinct NESSs. For \(\Omega < \Omega_c\), the system remains in a low-emission regime with negligible phonon back-action. In contrast, for \(\Omega > \Omega_c\), the emission power quickly increases and saturates at a much higher level, indicating enhanced triplon excitation and strong phonon-mediated amplification of the driven dynamics.
		
		In Fig.~\ref{f2}(c), for a slightly higher phonon frequency \(\omega_0/J = 1.35\), the transition persists but requires longer driving times. As the phonon approaches the two-triplon continuum, nonlinear feedback between phonons and triplons becomes more pronounced. The critical driving frequency shifts slightly downward to \(\Omega_c/J \approx 1.38\), reflecting increased detuning between the drive and the renormalized phonon mode.
		
		Thus, we address the questions posed in the introduction. We observe that the NESS evolves distinctly in the strongly hybridized phonon-triplon chain, exhibiting sudden jumps. To confirm that the observed jumps are sharp and correspond to first-order phase transitions, we present the response of additional phonon modes over a broader range of driving frequencies in two videos provided in the Supplemental Material (SM)~\cite{SM}: one showing phonons (\(\omega_0/J = 1.3\)--1.65) near or in the two-triplon band (with the transition), and the other showing phonons (\(\omega_0/J = 1.1\)--1.25) far from the two-triplon band (without the transition). The parameters used in the videos are \(g/J = 0.5\), \(g'/J' = 0.3\), \(\mathcal{E}/\omega_0 = 0.01\), \(\gamma_p/\omega_0 = 0.05\), and \(\gamma_s/J = 0.01\). The small red squares in the videos indicate the location of the first-order phase transition. From these videos, we notice that for driving frequencies \(\Omega \approx \omega_0\), the transient timescale of phonons far from \(2\varepsilon_0\) is governed solely by \(\gamma_p\), while for \(\Omega \approx 2\varepsilon_0\), it is influenced predominantly by \(\gamma_s\). However, for phonons close to \(2\varepsilon_0\), the transient timescale is governed by hybrid phonon-triplon dissipation arising from strong SPC. We also note that when the phonon lies within the two-triplon band, resonance and hybridization effects are maximal. In the NESS, this results in beating-type patterns with complex harmonic content. However, as the phonon penetrates deeper into the two-triplon band, this complex pattern disappears.

		\subsection{Analytical Floquet analysis in the NESS}
		Since the transitions emerge as a continuation of the transient dynamics—governed predominantly by dissipation—we attribute their origin to dissipative processes along with strong hybridization. This type of transition is dynamical in nature and does not rely on symmetry breaking. Instead, it arises from the nonlinear strong coupling between phonons and triplons, mediated by the periodic drive. Specifically, in the equation for \(\dot{n}_k(t)\) in Eq.~\eqref{eq_11c}, the term \(2 B_k q(t)\, \imaginary[z_k(t)]\) is nonlinear due to the product of \(q(t)\) and \(\imaginary[z_k(t)]\). Similarly, in the equation for \(\dot{z}_k(t)\) in Eq.~\eqref{eq_11d}, the terms \(2i A_k q(t) z_k(t)\) and \(2i B_k q(t)\left( n_k(t) + \tfrac{3}{2} \right)\) introduce additional nonlinearities. These terms indicate that the system can exhibit complex dynamical behavior such as bifurcations or instabilities. In what follows, we demonstrate that these nonlinearities manifest predominantly through the dissipation channel if the SPC is sufficiently strong.
		
	In the NESS, the mean-field oscillations can become synchronized with the drive, as observables exhibit periodic oscillations at an emergent frequency \(\omega\), which typically differs from both the bare phonon and drive frequencies. This frequency \(\omega\) arises from the combined effects of driving, interactions, and dissipation. Consequently, the mean-field observables can be expanded as a Fourier series, \( O(t) \approx \sum_m O_m e^{i m \omega t} \), leading to nonlinear mean-field equations for the components \(O_m\) of the Floquet-space vector representing $O(t)$. Nonetheless, due to the stability of the periodic oscillations under Lindblad dynamics, Floquet theory remains applicable. 
	Accordingly, we begin with \( m = 1 \) to explore the emergence of analytical solutions. However, our primary interest, as dictated by Eq.~\eqref{eq_12}, lies in the triplon density \( n_k(t) \). Given that the chain hosts \( 3(L + 1) \) triplon modes, we focus on the band-edge modes to simplify the analysis. This approximation is justified since the triplon density of states is enhanced at \( k = 0 \) and \( k = \pi \), where mode degeneracy amplifies nonlinear effects. Therefore, nonlinear instabilities are expected to be most pronounced at these points. In what follows, we focus specifically on the \( k = 0 \) mode. 
		
		For \( m = 1 \), Eqs.~\eqref{eq_11a} and~\eqref{eq_11b} become~(for simplicity, we denote $\varepsilon_{k = 0} = \varepsilon_0$, $A_{k=0} = A_0$, $B_{k=0} = B_0$, \( \overline{n}_{k=0} = \overline{n}\), \(\overline{z}_{k=0} = \overline{z} \), \(n_{1,k=0} = n_1\), and \(z_{1,k=0} = z_1\))\begin{subequations}
			\begin{align}
				i \omega q_1 = {} & \omega_0 p_1 - \tfrac{\gamma_p}{2} q_1\, ,\label{eq_13a}\\
				i \omega p_1 = {} & - \omega_0 q_1 - 2 \mathcal{E} - \tfrac{2 A_0}{L} n_1 - \tfrac{2 B_0}{L} \real[z_1] -  \tfrac{\gamma_p}{2} p_1\, .\label{eq_13b}
			\end{align}
		\end{subequations}Substituting \( p_1 = \frac{i \omega + \frac{\gamma_p}{2}}{\omega_0} q_1 \) from Eq.~\eqref{eq_13a} into Eq.~\eqref{eq_13b}, we obtain\begin{align}\label{eq_14}
			q_1 = \frac{2 \omega_0\left(\mathcal{E} + \tfrac{A_0}{L} n_1 + \tfrac{B_0}{L} \real[z_1] \right)}{\omega^2 - \omega_0^2 - \frac{\gamma_p^2}{4} - i \gamma_p \omega}\, .
		\end{align}Similarly, Eqs.~\eqref{eq_11c} and~\eqref{eq_11d} become\begin{subequations}
			\begin{align}
				i \omega n_1 = & 2 B_0 \,\overline{q}\, \imaginary[z_1] -\gamma_s n_1\, ,\label{eq_14a}\\
				i \omega \real[z_1]= & -2 \left(\varepsilon_0 + A_0\, \overline{q}\right) \imaginary[z_1] - \gamma_s\real[z_1]\, ,\label{eq_14b}\\
				i \omega \imaginary[z_1]= & 2 \left(\varepsilon_0 + A_0\, \overline{q}\right) \real[z_1] + 2 B_0 \,q_1 \left(\overline{n} + \tfrac{3}{2}\right) - \gamma_s\imaginary[z_1]\, ,\label{eq_14c}
			\end{align}
		\end{subequations}Substituting \( \imaginary[z_1] = -\frac{i \omega + \gamma_s}{2 \left(\varepsilon_0 + A_0\, \overline{q}\right)}  \real[z_1]\) from Eq.~\eqref{eq_14b} into Eq.~\eqref{eq_14c}, we obtain\begin{align}\label{eq_16}
			\real[z_1] = \frac{4 B_0\, q_1\,\left(\overline{n} + \tfrac{3}{2}\right)\left(\varepsilon_0 + A_0\, \overline{q}\right)}{\omega^2 -4 \left(\varepsilon_0 + A_0\, \overline{q}\right)^2 - \gamma_s^2 - 2 i \gamma_s \omega }\, ,
		\end{align}which gives rise to\begin{align}\label{eq_17}
			\imaginary[z_1] = \frac{-2 B_0\, q_1\,\left(\overline{n} + \tfrac{3}{2}\right)\left(i \omega + \gamma_s\right)}{\omega^2 -4 \left(\varepsilon_0 + A_0\, \overline{q}\right)^2 - \gamma_s^2 - 2 i \gamma_s \omega }\, .
		\end{align}These reflect the dynamical polarization due to phonon-triplon interaction. On the other hand, from Eq.~\eqref{eq_14a}, one can identify the first harmonic of the triplon density as\begin{align}\label{eq_18}
			n_1 = \frac{-4 B^2_0\, \overline{q}\, q_1\,\left(\overline{n} + \tfrac{3}{2}\right)}{\omega^2 -4 \left(\varepsilon_0 + A_0\, \overline{q}\right)^2 - \gamma_s^2 - 2 i \gamma_s \omega }\,. \end{align}As we need $\overline{q}$ in these relations, one can readily use Eqs.~\eqref{eq_11a} and~\eqref{eq_11b} for the zeroth harmonics to find\begin{align}\label{eq_19}
			\overline{q} = {} \frac{- 2 \omega_0 \left(A_0 \overline{n} + B_0 \overline{\real[z]}\right)}{L\left(\omega_0^2 + \tfrac{\gamma_p^2}{4}\right)}\, .
		\end{align}
		
		The system exhibits signatures of singular behavior in the denominators of key observables, primarily due to the dissipation rates \(\gamma_p\) and \(\gamma_s\), which contribute imaginary parts. We solve the equations self-consistently using pragmatic initial guesses to prevent divergences and to seed the nonlinear feedback loops that reveal resonance and singular behavior near the dynamical phase transition. The resulting solutions exhibit resonance signatures, nonlinearities, and singularities associated with instabilities and phase transitions, which appear in the solutions of the system’s order parameter, such as the triplon density \(\overline{n}\). When multiplied by \(\varepsilon_0\) and \(\gamma_s\), \(\overline{n}\) reflects the same features as the power described in Eq.~\eqref{eq_12}.
		
		In these relations, the matrix element \( B_0 = (g J' - g' J) / (2 \varepsilon_0) \) plays a crucial role in the dynamics. When the ratio condition \( g/J = g'/J' \) is met, \( B_0 \) vanishes, causing both \( n_1 \) and \( z_1 \) to vanish as well. In this limit, the phonon displacement reduces to \( q(t) = \overline{q} + q_1 \cos(\omega t) \). Assuming \( \gamma_p^2 / 4 \to 0 \), we find \( \overline{q} = -2 A_0 \overline{n} / (L \omega_0) \) and \( q_1 = 2 \omega_0 \mathcal{E} / (\omega^2 - \omega_0^2 - i \gamma_p \omega) \), which corresponds to the standard response of a driven-damped harmonic oscillator. This is the regime where the phonon-triplon system is effectively decoupled. Thereby, our previous assumption that \( g/J < g'/J' \) remains valid.
		
		To assess the validity of the Floquet analysis, we compute the time-averaged triplon emission power in the NESS by averaging \(\mathcal{P}(t)\) over one drive period as $\overline{\mathcal{P}} = \frac{\Omega}{2\pi} \int_0^{2\pi/\Omega} \mathcal{P}(t) \, dt$. The system is evolved for a sufficiently long time, approximately \( t J \approx 20000 \), to ensure convergence to the NESS before averaging. We then numerically evaluate \(\overline{\mathcal{P}}\) as a function of system parameters, especially the driving frequency \(\Omega\), and compare the results with the analytical Floquet predictions to identify the critical frequencies where the phase transitions occur.\begin{figure}[t]
			\centering
			\includegraphics[width=0.9\linewidth]{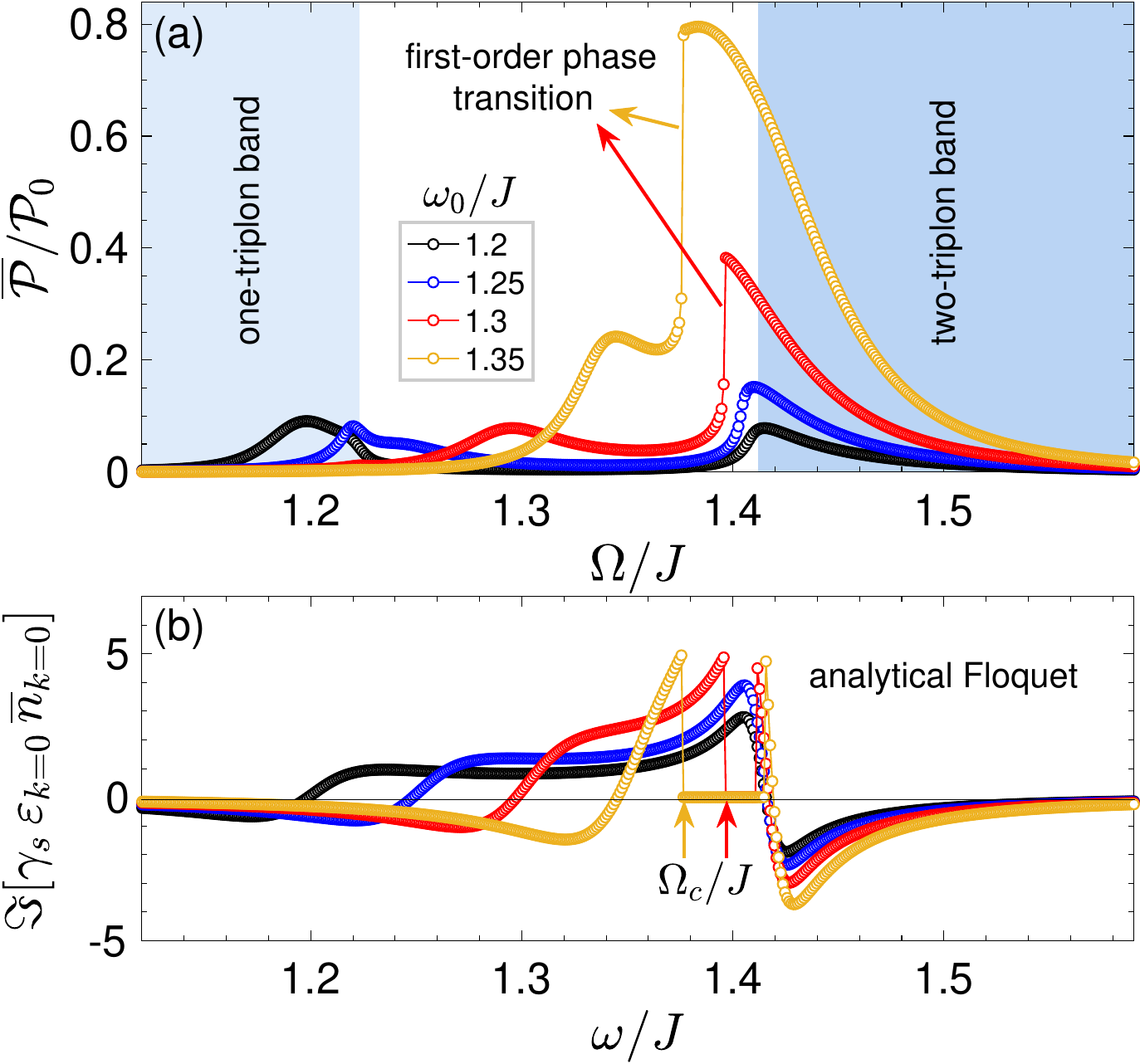}
			\caption{Dynamical first-order phase transition in the mean-field triplon emission power of a driven-dissipative dimerized spin-\(1/2\) chain. (a) Numerical time-averaged mean-field \( \overline{\mathcal{P}} \) in the NESS as a function of the driving frequency \( \Omega / J \) for various phonon frequencies \( \omega_0 / J \). (b) Analytical mean-field Floquet predictions for the imaginary part of \( \gamma_s\, \varepsilon_{k = 0} \overline{n}_{k=0} \) plotted versus the frequency \( \omega / J \) for the same phonon parameters, demonstrating excellent agreement at the critical driving frequencies. Parameters: \( g/J = 0.5 \), \( g'/J' = 0.3 \), \( \mathcal{E}/\omega_0 = 0.01 \), \( \gamma_p/\omega_0 = 0.05 \), and \( \gamma_s/J = 0.01 \).
			}
			\label{f3}
		\end{figure}
		
		Figure~\ref{f3}(a) shows the numerical results for the time-averaged emission power \(\overline{\mathcal{P}}\) as a function of the driving frequency \(\Omega/J\), computed with a step size of 0.001, for various phonon frequencies \(\omega_0/J\). Two of these phonon frequencies remain detuned from the lower edge of the two-triplon band, $2 \varepsilon_0$. Due to resonance effects, two prominent peaks are observed near \(\Omega \approx \omega_0\) and \(\Omega \approx 2 \varepsilon_0\). Additionally, a weaker sub-resonance peak linked to the one-triplon upper band edge appears, consistent with expected spectral features. For phonon frequencies \(\omega_0/J = 1.3\) and 1.35, these band-edge resonances become less distinct and merge into hybridized resonances, reflecting strong phonon-triplon coupling as the phonon mode approaches the two-triplon continuum. The sharp jumps in \(\overline{\mathcal{P}}\) at critical driving frequencies \(\Omega_c/J \approx 1.4\) and \(1.38\), respectively, mark the onset of a dynamical first-order phase transition in the NESS. 
		
		The analytical Floquet results in Fig.~\ref{f3}(b) show two zeros at \(\omega_1 \approx \omega_0\) and \(\omega_2 \approx 2 \varepsilon_0\) when the phonon mode is far detuned from the two-triplon band (for \(\omega_0/J = 1.2\) and \(1.25\)). However, an additional zero appears at \(\omega_3 \approx \sqrt{2(\varepsilon_0 + A_0 \overline{q})^2 - \gamma_s^2}\) once the phonon frequency approaches the two-triplon band (for \(\omega_0/J = 1.3\) and \(1.35\)), indicating strong hybridization effects. These analytical results align well with the numerical responses in Fig.~\ref{f3}(a), where resonance peaks occur near \(\omega_1\), and the first-order phase transition occurs at the critical driving frequency \(\Omega_c = \omega_3\). Thus, we confirm that the strongly hybridized phonon-triplon chain inherently exhibits first-order phase transitions in the nonequilibrium regime. These transitions arise when the system achieves a delicate balance between external driving forces and dissipation mechanisms, i.e., NESS, highlighting the crucial interplay between coherent drive and energy loss in stabilizing distinct dynamical phases.
		
		\subsection{Phase diagram and parameter sensitivity}
		
		Next, we systematically scan the parameter space as a function of the driving frequency to investigate the shift of the critical frequency \(\Omega_c\). Figure~\ref{f4} presents the triplon emission power in the NESS plotted versus both the driving frequency \(\Omega/J\) and the phonon frequency \(\omega_0/J\). A clear diagonal resonance line \(\Omega \approx \omega_0\) is observed, with slight deviations due to strong SPC effects and spectral broadening caused by phonon-triplon hybridization. Beyond this primary resonance, the phase diagram also exhibits features corresponding to the one- and two-triplon band edges across a wide parameter range. Building on observations from Fig.~\ref{f3}, we confirm both numerically and analytically that the phonon energy must be near the two-triplon continuum to trigger the nonlinear dynamics underlying the first-order phase transition. This nonlinear regime onset occurs around \(\omega_0/J \approx 1.27\), as indicated by the dotted red lines showing the analytical Floquet predictions, which are in excellent agreement with the numerically determined transition points.\begin{figure}[t]
			\centering
			\includegraphics[width=0.9\linewidth]{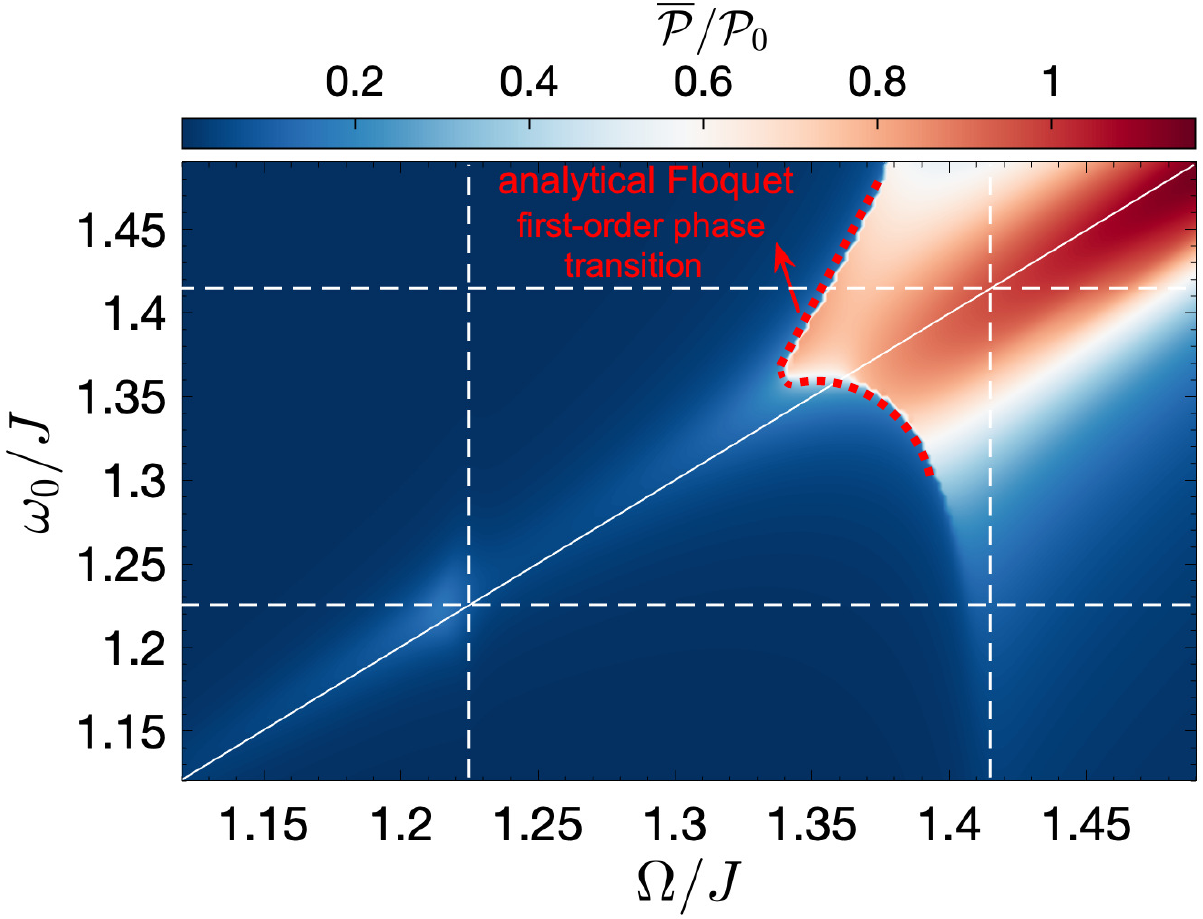}
			\caption{Mean-field phase diagram showing the triplon emission power as a function of driving frequency \(\Omega/J\) and phonon frequency \(\omega_0/J\). The diagonal line reflects resonance effects at \(\Omega \approx \omega_0\), with slight deviations due to strong SPC and hybridization. The diagram also captures signatures of the one- and two-triplon band edges over a wide parameter range. For \(\omega_0/J < 1.27\), where the phonon energy lies well below the two-triplon continuum, only modest shifts in the emission edge occur as \(\Omega\) varies. In contrast, for \(\omega_0/J > 1.27\), sharp, step-like features appear in the emission power, signaling first-order phase transitions. As \(\omega_0/J\) approaches \(\sqrt{2}\), the critical driving frequency \(\Omega_c\) decreases, reaching a minimum near \(\omega_0/J \approx 1.37\), before rising again. This non-monotonic trend arises from an optimal overlap between phonon and laser resonances combined with an instability. The dotted red line traces the analytical Floquet prediction from Eqs.~\eqref{eq_14}--\eqref{eq_19}, showing excellent agreement with the numerically determined transition points. Parameters: \(g/J = 0.5\), \(g'/J' = 0.3\), \(\mathcal{E}/\omega_0 = 0.01\), \(\gamma_p/\omega_0 = 0.05\), and \(\gamma_s/J = 0.01\).
			}
			\label{f4}
		\end{figure}\begin{figure*}[t]
			\centering
			\includegraphics[width=1\linewidth]{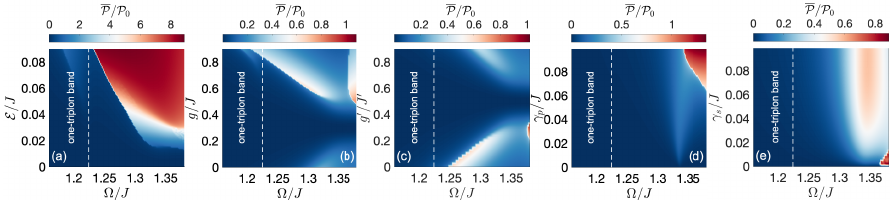}
			\caption{Parameter sensitivity around a known $\Omega_c/J$ in the mean-field triplon emission power \(\overline{\mathcal{P}}\) for phonon energy \(\omega_0/J = 1.35\). Panel (a) shows the response versus laser amplitude \(\mathcal{E}/J\); panels (b)–(c) illustrate the effects of SPC strengths \(g/J\) and \(g'/J'\), respectively; panels (d)–(e) depict the influence of phonon and spin damping rates \(\gamma_p/J\) and \(\gamma_s/J\). The vertical dashed line indicates the upper boundary of the one-triplon band. The nick, step-like changes in emission power signal first-order phase transitions, revealing strong nonlinear dynamics and parameter sensitivity in the driven-dissipative chain. In panel (a), we fix the SPCs at $g/J = 0.5$ and $g'/J' = 0.3$. In panel (b), we fix the laser amplitude to $\mathcal{E}/J = 0.0135$ and set $g'/J' = 0.3$, while in panel (c), the same laser amplitude is used with $g/J = 0.5$. For panels (a)–(c), the damping rates are fixed at $\gamma_p/J = 0.0675$ and $\gamma_s/J = 0.01$. In panel (d), we fix $\mathcal{E}/J = 0.0135$, $g/J = 0.5$, $g'/J' = 0.3$, and $\gamma_s/J = 0.01$, while in panel (e), the same values for $\mathcal{E}$, $g$, and $g'$ are used with $\gamma_p/J = 0.0675$.}
			\label{f5}
		\end{figure*} 
		
		The phase boundaries indicate a discontinuous change in the system’s order parameter—specifically, the triplon density, which is directly reflected in the emission power. The steep gradients near these boundaries highlight a strong nonlinear response arising from the hybridization between lattice phonons and triplon excitations. Outside these regions, the system supports coherent triplon dynamics, potentially stabilized by avoided crossings or hybridization with phonon modes. Overall, within the range \( 1.25 < \omega_0/J < 1.5 \), the first-order phase transition occurs over a narrow window of critical driving frequencies, \( 1.3 < \Omega_c/J < 1.4 \). As the phonon mode penetrates deeper into the two-triplon band, the transition weakens and is expected to vanish for \( \omega_0/J > 1.7 \), as shown in the videos provided in the SM~\cite{SM}.
		
		Furthermore, a non-monotonic behavior emerges near the point \( (\omega_0, \Omega_c) \approx (1.37, 1.34) \), where the critical driving frequency \(\Omega_c\) initially decreases with increasing phonon frequency \(\omega_0\), but then reverses its trend beyond \(\omega_0 \approx 1.37\). The phase boundary—separating the phonon-dominated and laser-driven regimes—exhibits a dip at this point before rising again. This feature results from the optimal overlap between phonon and laser resonances intersecting with an instability near this region. Following this intersection, the system displays a precursor-like response, signaling a transition into a new monotonic regime with the opposite trend.
		
		To further address the questions posed in the introduction, we analyze the parameter sensitivity of $\Omega_c$. Given the tightly coupled equations of motion and the strong nonlinearities in the system's response, it is plausible that multiple critical driving frequencies may trigger phase transitions. Although there are various choices for the initial parameters, to explore this possibility, we focus on the phonon frequency \(\omega_0/J = 1.35\), where the critical driving frequency is known to occur near \(\Omega_c/J \approx 1.38\) for the chosen parameter set. Accordingly, we fix the upper bound of the driving frequency at 1.38 and vary other system parameters to see how the critical point shifts.
		
		The variation of the emission power with laser amplitude \(\mathcal{E}/J\) in Fig.~\ref{f5}(a) reveals a pronounced onset of high emission beyond a threshold amplitude of approximately \(0.03\,J\). This behavior aligns with the role of the laser field as a driving perturbation, which naturally enhances energy absorption and emission as the field strength increases~\cite{PhysRevB.103.045132,PhysRevB.110.064420}. Notably, starting from the critical frequency \(\Omega_c/J \approx 1.38\), the transition line shifts toward lower driving frequencies with increasing laser amplitude, indicating that stronger fields destabilize the low-emission phase at progressively lower frequencies. This trend suggests an enhanced susceptibility of the system to the external drive under stronger excitation. Furthermore, the emergence of features below the one-triplon band edge at high fluence (bright area) reflects a dynamical modification of the band structure induced by the strong drive. This shift is mediated by the spin-phonon interaction, which modulates the excitation spectrum in a nonlinear and fluence-dependent manner.
		
		In Fig.~\ref{f5}(b), we track the dependence of the emission power on the SPC strength along the \(J\) bond, \(g/J\). As \(g/J\) increases beyond 0.5, the overall emission power decreases slightly, and the original first-order phase transition near the critical driving frequency \(\Omega_c/J \approx 1.38\) shifts modestly. Remarkably, for \(g/J > 0.5\), a second first-order phase transition emerges at lower driving frequencies, shifting linearly and monotonically toward the one-triplon band edge as \(g/J\) increases. This behavior reveals the coexistence of two distinct dynamical first-order phase transitions. The appearance of this additional transition suggests that stronger SPC enhances the nonlinear feedback between spin dynamics and lattice distortions, thereby promoting an earlier onset of the transition.
		
		A similar analysis for \( g'/J' \) (the alternate SPC on the \( J' \) bond) in Fig.~\ref{f5}(c) reveals a less pronounced sensitivity near the critical driving frequency \(\Omega_c/J \approx 1.38\), exhibiting step-like modulations in the emission power. Nevertheless, the emergence of a second first-order phase transition is favored for \( g'/J' < 0.3 \). In contrast to the behavior observed for \( g \), where two transitions coexist over a broad range \( g/J > 0.5 \), the coexistence regime for \( g' \) is confined to a narrow window \( 0.2 < g'/J' < 0.25 \). For \( g'/J' < 0.2 \), only the second transition persists, occurring at lower driving frequencies. These discrete features indicate a multi-stage control mechanism in the strong hybridization regime, influenced by the dissipation channel within the nonlinear dynamical landscape.
		
		We now focus on the role of damping, as illustrated in Figs.~\ref{f5}(d) and~\ref{f5}(e), which represents a crucial element of the dynamical landscape in the strongly hybridized regime. Notably, in the absence of both spin and phonon damping, no signatures of a phase transition are observed, and no phase boundary can be defined. As the phonon damping rate \(\gamma_p/J\) increases, the system remains in the low-emission regime over a broader range of driving frequencies \(\Omega\), effectively delaying the transition to the high-emission phase. For phonon damping strengths \(\gamma_p/J > 0.06\), the critical driving frequency shifts toward lower values; however, no evidence of a second first-order phase transition emerges. Spin damping \(\gamma_s/J\), on the other hand, causes a pronounced delay and broadening of the transition. At large \(\gamma_s\), the emission remains suppressed even beyond the upper phonon edge, indicating that spin relaxation can effectively inhibit the dynamical instability responsible for the phase transition. Interestingly, even a very small but finite spin damping induces a sharp jump in the emission power, provided the phonon damping exceeds approximately 6\% of the magnetic exchange \(J\). This analysis confirms that, in the absence of dissipation, the system’s nonlinearities alone from strong hybridization are insufficient to trigger phase transitions.
		
		Overall, the location and character of the phase transitions are strongly influenced by the interplay between periodic driving, phonon-triplon coupling strength, and dissipation. Among these factors, dissipation plays a particularly crucial role, as it governs the onset and structure of the nonlinearities that ultimately determine the nature of the steady-state phases. Although the detailed phase landscape may shift depending on the phonon frequency—especially when it lies within or near the edge of the two-triplon continuum—the qualitative behavior remains robust. Notably, such scenarios can be systematically analyzed using the Floquet-based framework and numerical approach developed in this work, enabling a straightforward extension of our results to neighboring parameter regimes.
		
		\subsection{Experimental perspective}
		From an experimental perspective, the observed first-order phase transitions—arising from the strong hybridization of phonon and triplon excitations near the two-triplon continuum in the presence of dissipation—can be verified by measuring the emission power spectrum using time-resolved THz pump-probe techniques~\cite{PhysRevB.100.165131,PhysRevLett.124.057404,doi:10.1073/pnas.2204219119,doi:10.1021/acs.jpcc.0c06344,PhysRevLett.107.107401,PhysRevB.71.235110}. Time-resolved changes in absorption or reflectivity, which are related to the emission power, especially around two-triplon energy, should monitor the transitions. 
		
		Particular attention should be given to the driving frequency \(\Omega\) around \(1.35\text{--}1.4\,J\), which corresponds roughly to 3 THz to 3.5 THz for \(J \approx 10\,\mathrm{meV} \approx 2.42\,\mathrm{THz}\) in CuGeO$_3$. Accurate tuning of the laser amplitude to a fairly weak regime \((\mathcal{E} \approx 0.05\) THz) and exciting the infrared-active optical phonon with frequency \((\omega_0/J)\) within the same 3 THz to 3.5 THz range~($A_u$ and $B_{3u}$ modes~\cite{spitz2025phononspectrumspinpeierlsphase}) is crucial; this can be achieved using modern tunable THz sources. Additionally, the damping rates \((\gamma_p/J, \gamma_s/J) \approx (0.15, 0.025)\,\mathrm{THz}\) can be adjusted through control of the sample temperature or coupling to a high-quality bath sink~\cite{PhysRevB.103.045132}. Crucially, these parameter choices suppress heating by design, as the drive strength and dissipation rates are finely tuned to maintain a stable NESS.
		
		\section{Summary and outlook}\label{S4}
		Employing the mean-field Lindblad formalism, we reveal a dynamical first-order phase transition in hybrid phonon-triplon systems, driven by strong nonlinear feedback and dissipation effects. This transition prominently occurs when phonon frequencies lie near the two-triplon band in the strong spin-phonon coupling regime and is fundamentally governed by dissipation-induced nonlinearities. The critical driving frequency at which the transition emerges depends sensitively on coupling strengths, damping rates, and drive amplitude. Our mean-field analytical approach, based on the dominant Floquet steady-state harmonics, corroborates the numerical findings, providing a comprehensive understanding of the underlying mechanisms. 
		
		These results uncover rich dynamical phenomena in driven-dissipative strongly hybridized quantum systems and open promising pathways for advanced quantum control and nonlinear optical functionalities. They further motivate experimental efforts to probe and harness such transitions in real materials and engineered platforms.
		
		Although this study focused on harmonic phononic effects, the observed asymmetry in the phase diagrams and the gradual broadening of the triplon band with increasing emission power hint at additional higher-order phenomena, such as anharmonic phonon contributions and multi-phonon processes. Although nonlinear phononic effects can be incorporated theoretically, isolating and characterizing them experimentally remains challenging. This will likely require advanced spectroscopic techniques, such as resonant inelastic X-ray scattering, to fully resolve the excitation spectrum. 
		
		Furthermore, dynamically tuning the driving frequency—such as via frequency chirping—provides a novel way to control the phase transition by probing nonequilibrium effects beyond static conditions. By adiabatically preparing the system in a NESS above the critical frequency and then slowly sweeping downward, one may observe hysteresis with a shifted critical point, reflecting memory and nonlinear feedback~\cite{PhysRevB.109.224417,PhysRevB.108.L140305}. This hysteretic behavior enriches our understanding of dynamical phase transitions and offers new avenues for controlling quantum phases through time-dependent driving. We leave these promising directions for future work.
		
		\section*{Acknowledgments}
		M. Y. gratefully acknowledges Michael Kolodrubetz for insightful discussions. M. Y. was supported by the Department of Energy, Office of Basic Energy Sciences, Division of Materials Sciences and Engineering under Contract No. DE-FG02-08ER46542 for the formal developments, the analytical/numerical work, and the writing of the manuscript. 
	}
	\twocolumngrid
	\bibliography{bibliography.bib}
	
\end{document}